\documentclass[%
 reprint,
superscriptaddress,
 amsmath,amssymb,
 aps,
]{revtex4-1}
\usepackage{subfig}
\usepackage{amsmath}
\usepackage{graphicx}
\usepackage{epstopdf, epsfig}
\usepackage{bm}
\usepackage{color,soul}

\makeatletter
\newcommand{\thickhline}{%
    \noalign {\ifnum 0=`}\fi \hrule height 1pt
    \futurelet \reserved@a \@xhline
}
\begin{document}

\preprint{APS/123-QED}

\title{Droplet impact of Newtonian fluids and blood on simple fabrics: effect of fabric pore size and underlying substrate.}% Force line breaks with \\
%\thanks{A footnote to the article title}%

\author{T. C. de Goede}
\email{T.CdeGoede@uva.nl}
\affiliation{
 Van der Waals-Zeeman Institute, Institute of Physics, University of Amsterdam, Science Park 904, 1098 XH Amsterdam, Netherlands\\
}
\author{A. M. Moqaddam}
\affiliation{
Chair of Building Physics, Department of Mechanical and Process Engineering, ETH Zurich,
8092 Zurich, Switzerland\\}
\affiliation{Laboratory for Multiscale Studies in Building Physics, Empa, Swiss Federal Laboratories for Materials Science and Technology, 8600 Dubendorf, Switzerland\\}

\author{K. C. M. Limpens}
\affiliation{
 Van der Waals-Zeeman Institute, Institute of Physics, University of Amsterdam, Science Park 904, 1098 XH Amsterdam, Netherlands\\ }

\author{S. A. Kooij}
\affiliation{
 Van der Waals-Zeeman Institute, Institute of Physics, University of Amsterdam, Science Park 904, 1098 XH Amsterdam, Netherlands\\ }

\author{D. Derome}
\affiliation{Department of Civil and Building Engineering, Universit\'e de Sherbrooke, 2500, boul. de l'Universit\'e Sherbrooke (Qu\'ebec), Canada\\}

\author{J.Carmeliet}
\affiliation{
Chair of Building Physics, Department of Mechanical and Process Engineering, ETH Zurich,
8092 Zurich, Switzerland\\}

\author{N. Shahidzadeh}
\affiliation{
 Van der Waals-Zeeman Institute, Institute of Physics, University of Amsterdam, Science Park 904, 1098 XH Amsterdam, Netherlands\\ 
}

\author{D. Bonn}
 \email{d.bonn@uva.nl}
\affiliation{
 Van der Waals-Zeeman Institute, Institute of Physics, University of Amsterdam, Science Park 904, 1098 XH Amsterdam, Netherlands\\ 
}%
\date{\today}% It is always \today, today,
             %  but any date may be explicitly specified

\begin{abstract}
When a droplet impacts a fabric mesh at a sufficiently high impact velocity, it not only spreads over the fabric but also penetrate its pores. To determine the influence of this liquid penetration of the fabric on droplet spreading on thin fabric meshes, we measured the droplet spreading ratio on fabric with and without an underlying substrate using a high-speed camera. For fabrics without a substrate, the droplet spreading ratio is reduced as the fabric penetration by the liquid reduces the droplet volume spreading on top of the fabric. Using entropic lattice Boltzmann simulations, we find that the lower droplet spreading ratio on fabrics, both with and without a substrate, is due to an increase of viscous losses inside the droplet during spreading. Comparing droplet impact of blood with its Newtonian counterpart, we show that for spreading on fabrics, just like on smooth surfaces, blood can be approximated as a Newtonian fluid.  
\end{abstract}

%\pacs{Valid PACS appear here}% PACS, the Physics and Astronomy
                             % Classification Scheme.
%\keywords{Suggested keywords}%Use showkeys class option if keyword
                              %display desired
\maketitle

\section{Introduction}
\begin{figure*}
\centering
\includegraphics[width=.8\textwidth]{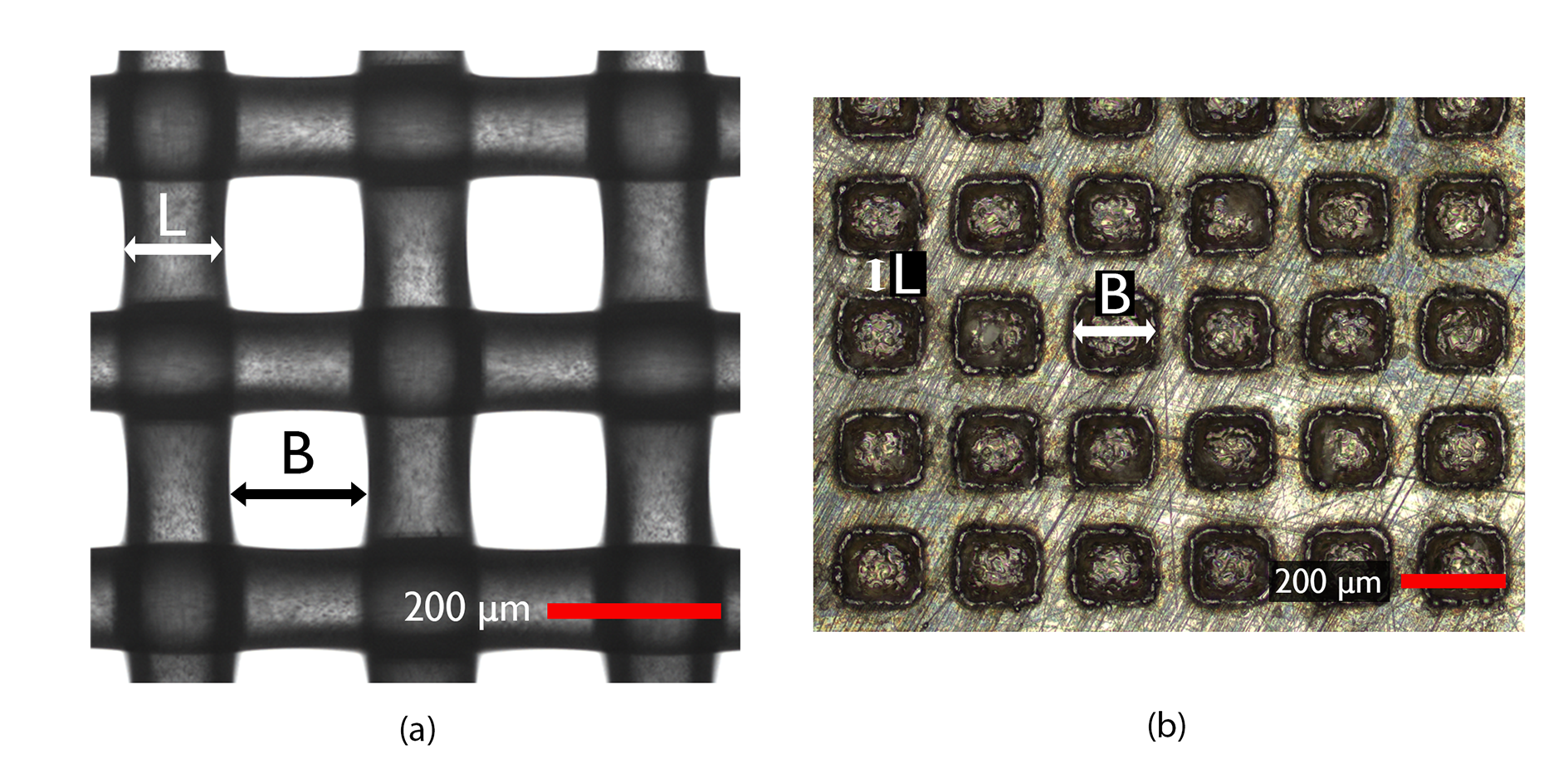}
\caption{(a) Microscope image of a plain woven monofilament polyester fabric. (b) Optical image (top) and height profile (bottom) of a patterned surface as measured with an optical laser microscope (Keyence VK-X1000). The average measured depth of the holes is equal to $64 \pm 1$ micron. Pore size $B$ and pore spacing (or yarn diameter) $L$ are denoted in both figures.}
\label{fig:methodexp}
\end{figure*}

Although the impact of a droplet onto a surface is a commonly occurring phenomenon, it has proven to be an interesting field of study due to the intricate interplay between inertial, capillary and viscous forces inside the droplet during impact. Studies have shown that a wide range of interesting physical phenomena during impact can occur: the droplet can simply spread over the surface \citep{Laan2014,Lee2015,Wildeman2016,deGoede2019}, but small satellite droplets can also detach from the droplet due to the interaction with the surrounding air (splashing) at high impact velocities \citep{Riboux2014,deGoede2017,Quetzeri2019}. Droplets can even completely bounce off the surface \citep{Kolinski2014drops,deRuiter2015,deRuiter2015b}. These phenomena depend not only on fluid parameters such as viscosity or surface tension, but also on the atmospheric conditions of the surrounding gas \citep{Xu2005,San2012,Kolinski2012,Kolinski2014,Sprittles2015,Sprittles2017} and surface properties such as wettability \citep{deGoede2019} and surface roughness \citep{Range1998}.\\

During impact, a droplet with initial diameter $D_0$ hits a surface at an impact velocity $v$,  spreading out until it reaches a maximum spreading diameter $D_{max}$. Recent studies \citep{Eggers2010,Laan2014,Lee2015} have established the relationship between droplet spreading ratio $D_{max} / D_0$ and density $\rho$, shear rate viscosity $\eta$ and surface tension $\sigma$ of the fluid by interpolating between two droplet spreading regimes where the kinetic energy of the droplet is either fully transformed into surface energy (capillary regime; $D_{max}/D_0 \propto \mathrm{We}^{1/2}$) \citep{Collings1990} or fully dissipated by the viscous forces inside the droplet (viscous regime; $D_{max}/D_0 \propto \mathrm{Re}^{1/5}$) \citep{Madejski1976,Roisman2002}. Here, We and Re are the Weber (We $\equiv \rho D_0 v^2/{\sigma}$) and Reynolds (Re $\equiv \rho D_0 v/\eta$) number, respectively. For low impact velocities, the wettability of the surface can also be incorporated into the droplet spreading model by using the spherical cap model \citep{Berthier2007,Lee2015,deGoede2019}.\\ 

While droplets spreading on smooth surfaces has received ample attention, droplet spreading on fabrics has not yet been studied extensively, despite its relevance to fields such as crime scene investigation and the textile industry. The fabric substrate complicates the physical picture. For example, absorption of liquid by the yarns of the fabric due to capillary action (wicking) \citep{Kissa1996,Nyoni2006,Benltoufa2008} could heavily influence the long timescale dynamics of droplet spreading. But even if the droplet is not absorbed by the fabric yarns, droplet spreading is still significantly different compared to droplet spreading on smooth surfaces. For example: recent studies on monofilament (fabric) meshes have shown that droplets penetrate through the pores of the mesh if their impact velocity is high enough, allowing a part of the droplets to pass through the mesh \citep{Brunet2009,Ryu2017,Soto2018,Kumar2018,Zhang2018,Kooij2019}. Several studies also showed that, at high Weber numbers, from the moment the fabric or mesh is penetrated, the droplet spreading ratio on fabrics is significantly lower than expected from the scaling models used in these studies \citep{Kumar2018,Zhang2018}. Understanding how droplet spreading is influenced by the liquid penetration of the (fabric) mesh and whether the pre-existing spreading model of Laan \textit{et al.} and Lee \textit{et al.} could be used to predict droplet spreading on these fabrics could have significant practical applications in for example the textile industry \citep{Ujiie2006,Fangueiro2010} and forensic research \citep{Williams2016,Castro2015,Li2017}. For forensic applications, the liquid of most interest is blood. As a result, it is also important to determine whether the shear thinning properties of blood have an influence on droplet spreading.
For droplet impact on smooth surfaces, \cite{Laan2014} have shown that the high shear rate exerted on a droplet during spreading allows blood to be approximated as a Newtonian fluid during droplet impact, with a viscosity given by the high-shear-rate viscosity $\eta_\infty$. In this specific case, the question is whether the same approximation holds for droplets spreading over fabric meshes, or that the liquid penetration of the fabric might lead to some shear thinning effects that do not occur on a smooth surface. A final question inspired by real-world experimental scenarios is how a solid substrate located underneath a fabric would influence the droplet spreading process.\\

In this study, we investigate the influence of liquid penetration of fabrics on droplet spreading for plain woven monofilament polyester meshes. Using high-speed imaging, we show that the aforementioned decrease in droplet spreading can be ascribed to part of the volume of the droplet passing through the fabric, in agreement with earlier studies. We find that, even if the downward flow of liquid beyond the fabric mesh is blocked by an underlying substrate, at high impact velocities ($v > 1$ m/s), the droplet spreading ratio is still lower than that of smooth surfaces. This difference is due to an increased viscous dissipation inside the droplet during spreading, as we conclude from combining experimental results and entropic lattice Boltzmann simulations. These simulations indicate that this increase in viscous dissipation is due to liquid flow into the pores of the fabric and to the droplet pushing itself between the fabric and substrate at velocities above the penetration velocity. We find that the shear thinning properties influence fabric penetration but not droplet spreading on top of the fabric, allowing blood to be approximated as a Newtonian fluid when considering droplet spreading on top of thin fabric meshes.

\section{Methods}
\begin{figure*}
	\centering
	\includegraphics[width=0.6\textwidth]{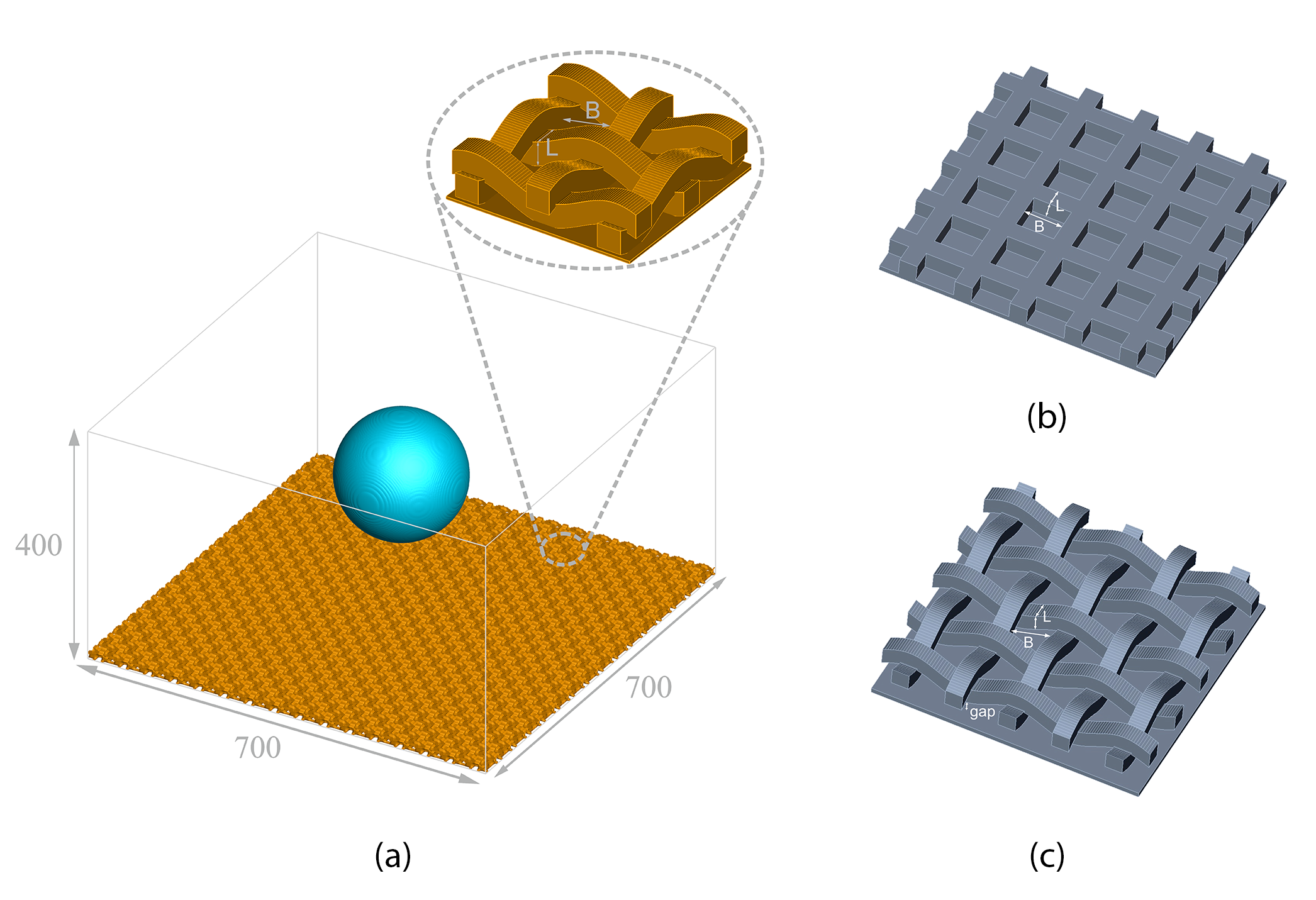}
	\caption{(a) Schematic representation of the simulation setup showing a droplet above a fabric-like geometry before impact. Droplet impact was simulated on for geometries representing a smooth surface (not shown), a patterned surface (b) and fabric (c). The pore size and yarn diameter for the latter two cases correspond to the 150-micron fabric investigated experimentally (Table \ref{tab:fabparm}). The fabric (c) was either attached to a substrate or placed above a substrate with a small gap of around 35 micron.}
	\label{fig:Schematic}
\end{figure*}

\begin{figure*}
\centering
\includegraphics[width=\textwidth]{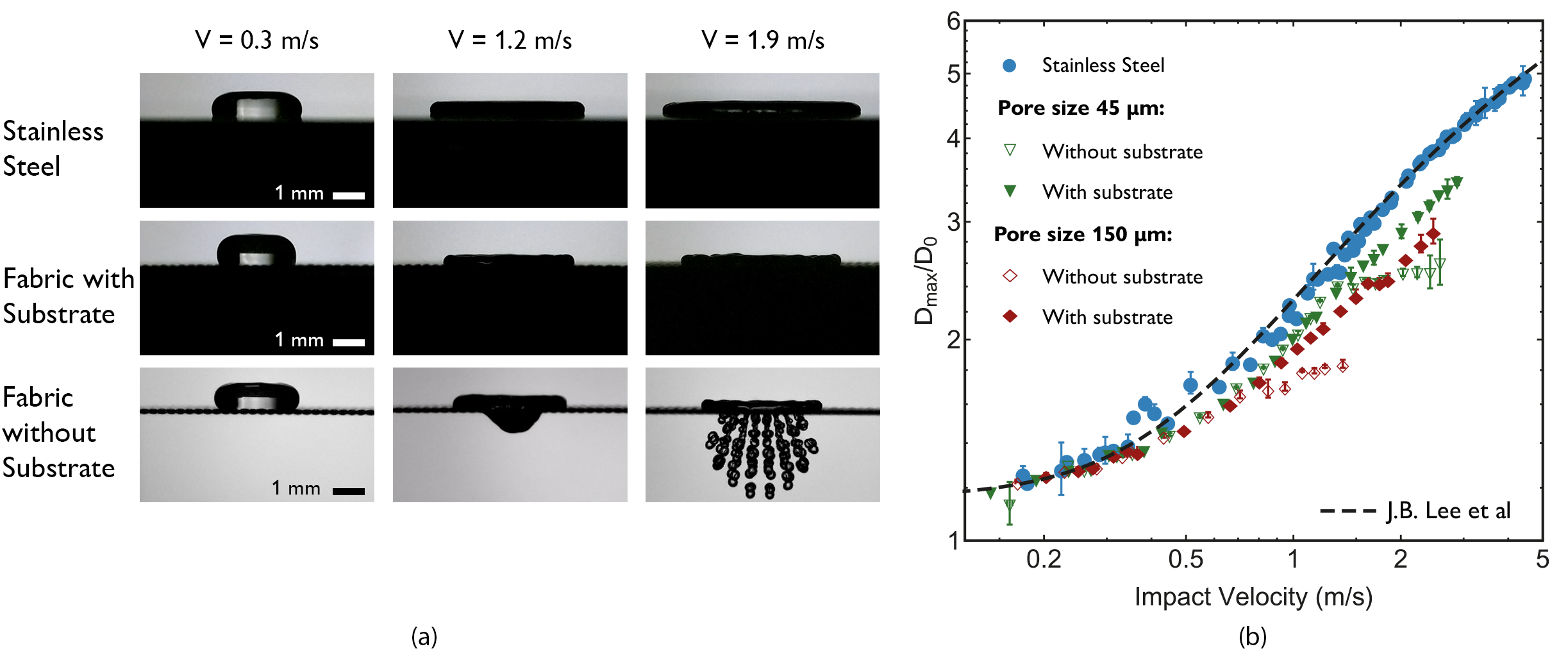}
\caption{(a) High-speed images of droplets at maximum spreading on stainless steel (top), fabric with substrate (middle) and fabric without substrate (bottom) at different impact velocities, increasing from left to right. The pore size of the fabric depicted here is 150 $\mu$m. (b) Measured spreading ratio as function of the impact velocity for fabrics with pore sizes of 150 micron (red symbols) and 45 micron (green symbols) compared to the droplet spreading ratio on stainless steel (blue circles), and the theoretical prediction of Lee \textit{et al.} (Eq. \ref{eq:spreadingmodel}) for droplet spreading on smooth surfaces (black dashed line). The filled and open symbols indicate whether the fabric was placed on a substrate or suspended in the air, respectively.}
\label{fig:highspeedimg}
\end{figure*}

\subsection{Experiments}

\begin{table}
\centering
\renewcommand{\arraystretch}{1.1}
\begin{tabular}{l c c }
System & Pore size $B$ ($\mu$m) & Pore spacing $L$ ($\mu$m)\\ 
\rule{0pt}{10pt}Fabric  & 45                 & 40\\ 
Fabric & 106				   & 70\\
Fabric & 150             & 80\\ 
Patterned surface & 135                & 76 \\ 
\end{tabular}
\caption{Pore size and spacing of the fabrics and patterned surface used.}
\label{tab:fabparm}
\end{table}

To measure the spreading ratio of an impacting drop, a 0.4 mm-diameter blunt-tipped needle was used to generate droplets with an initial diameter $D_0$ of around 2.3 mm falling from a variable height onto a substrate. The generated droplets were spherical during free fall, indicating that the droplets were sufficiently smaller than the capillary length of a water drop (2.7 mm) for gravity effects to be neglected in this study. By systematically changing the height of the needle, the maximum spreading diameter $D_{max}$ was investigated as a function of the impact velocity $v_{imp}$ using high-speed imaging with a frame rate between 4004-8100 fps and spatial resolutions between 11 and 17.7 micron per pixel. Droplet impact measurements were performed on a stainless steel surface as reference (average roughness $R_a = 0.247 \pm 0.007 \hspace{0.1cm} \mu$m as measured by a Keynece VK-X1000 laser microscope). As fabrics, single cylindrical polyester fibres were woven in a crisscross pattern, creating a polyester mesh with rectangular pores (Fig. \ref{fig:methodexp}a; Gilson Company Inc.). These single fibre yarn fabrics were chosen to eliminate any influence of liquid imbibition by the fabric on droplet spreading. Three different monofilament polyester fabrics were used, each with a different pore size and yarn thickness (see Table \ref{tab:fabparm}). For the polyester fabrics, droplet impact was measured on fabrics that were either spanned over a small gap of roughly 8 mm (fabric without substrate) or placed on a steel substrate (fabric with substrate). The fabric was spanned tight both over the gap and on the substrate to minimise unwanted energy loss due to fabric movement \citep{Chantelot2018,Kumar2018} or any influence of the tension exerted on the fabric \citep{Kooij2019}. In this study, it was assumed that the polyester fibres were rigid and do not deform during droplet impact. Placing the fabric on a substrate blocks the downward liquid flow when the fabric is penetrated, allowing us to investigate the effect of droplet penetration on droplet spreading. Droplet impact was also compared between plasma-treated fabrics and untreated fabrics to investigate the influence of fabric wettability on droplet spreading. Fabrics were plasma treated for 6 minutes on both sides to guarantee that the wettability was increased evenly across the whole fabric.\\

Droplet spreading on the fabrics with substrate was also compared to droplet spreading on a patterned surface: a steel surface in which rectangular holes were cut using an electrical discharge machining method \citep{Ho2003} (Fig. \ref{fig:methodexp}b). The size, depth and spacing of the holes were chosen in such a way that the dimensions of the pores of the patterned surface were similar to that of the 150-micron fabric (see Table \ref{tab:fabparm}).\\

For most experiments, demineralised water was used (density $\rho = 998$ kg/m$^3$, surface tension $\sigma = 72$ mN/m and viscosity $\eta = 1 $mPa s). Droplet impact of a 1:1 water-glycerol mixture ($\rho = 1124$ kg/m$^3$, $\sigma = 65.75$ mN/m and $\eta = 4$ mPas) was measured on both a smooth surface and 150-micron fabric. The water-glycerol measurements on a smooth surface were compared to droplet spreading of water on fabric, while the water-glycerol mixture measurements on the 150-micron fabric were compared with blood spreading on fabric. These two measurements were compared to investigate the influence of the shear thinning properties of blood on droplet spreading.  The properties of blood ($\rho = 1055$ kg/m$^3$, $ \sigma = 59$ mN/m and shear rate viscosity $\eta_{\infty} = 4.8$ mPas) were obtained from \cite{Laan2014}. Sodium citrate was added to blood to prevent coagulation during the experiments. The surface tension of blood was found to increase with roughly $3$ mN/m for a sodium citrate concentration of 10\%. Since the amount of sodium citrate that was added to the blood in this study was small ($\sim 2.5 \%$ mass concentration), we assume the change in surface tension due to the anti-coagulant to be negligible.\\

\subsection{Simulations}

For numerical simulations, the recently tested and validated entropic lattice Boltzmann method for two-phase flows was used \citep{ref12}. The numerical implementation and validation was discussed in more details elsewhere \citep{ref13,ref14,ref21,ref22,ref17,ref18}. Below we briefly describe the method together with a short clarification on the numerical set-up and simulated parameters used in this study.\\

For a liquid-vapour system separated by an interface, the entropic lattice Boltzmann equation reads as: 
\begin{equation}
\begin{split}
f_i(\bm{x}+\bm{v}_i\delta t,t+\delta t)= f_i(\bm{x},t)+
\alpha \beta\left[f_i^{\rm eq}(\rho,\bm{u})-f_i(\bm{x},t)\right]+ \\
[f_i^{\rm eq}(\rho,\bm{u}+\delta \bm{u}) - f_i^{\rm eq}(\rho,\bm{u})], 
\label{eq:Collision}
\end{split}
\end{equation}

where $f_i (\bm x,t)$ are the discrete populations and $v_i$ $ (i=1,…,N)$ denote the discrete velocities corresponding to the underlying lattice structure. The $\rm {D3Q27}$ lattice $(N=27)$ was used for our three-dimensional simulations. Parameter $\beta$ $(0<\beta<1)$ was determined using the kinematic viscosity, $\nu=c_{\rm s}^2\delta t[1/(2\beta)-1/2]$ with $c_s=\delta x/(\sqrt{3}\delta t)$ the lattice speed of sound and $\delta x=\delta t=1$ as lattice units. The equilibrium population $f_i^{eq}$ was used as the minimizer of the discrete entropy function $H=\sum_{i=1}^N f_i\ln(f_i/W_i)$, under the constraints of local mass and momentum conservations, $\{\rho,\rho \bm{u}\}=\sum_{i=1}^N\{1,\bm{v}_i\}\{f_i^{\rm eq}\}$, where $W_i$ are the lattice weights. The stabilizer parameter $\alpha$ defines the maximal over-relaxation, which is computed from the entropy estimate equation at each time step for each computational node  \citep{ref14}. \\

In Eq. \ref{eq:Collision}, the two-phase effects resulting from intermolecular forces are present through the velocity increment $\delta\bm{u}=(\bm{F}/\rho)\delta t$, with the force $\bm{F}$ being the sum of the fluid-fluid ($\bm{F}_{\rm {f-f}}$) and fluid-solid interactions ($\bm{F}_{\rm {f-s}}$). Phase separation occurs by defining the fluid-fluid interaction as $\bm{F}_{\rm {f-f}}=\nabla\cdot \left(\rho c_{\rm s}^2 \bm{I}-\bm{P}\right)$ using the Korteweg’s stress $\bm P$ as,
\begin{equation}
\bm{P}= \left(p-\kappa \rho \nabla ^2\rho-\frac{\kappa}{2}\left|\nabla \rho \right|^2\right)\bm{I}+\kappa (\nabla\rho) \otimes (\nabla\rho),
\label{eq:Pabtarget}
\end{equation}

where $\kappa$ is the coefficient controlling the surface tension, $\bm I$ the unit tensor and $p$ denotes the non-ideal equation of state \citep{ref15}, for which the Peng-Rabinson equation was used \citep{ref16}. The introduction of a cohesive interaction through the velocity increment in Eq. \ref{eq:Collision} leads to the surface tension forces separating the liquid and vapour by an interface, which maintains the liquid and vapour in an equilibrium state. The wettability condition is modelled by taking into account the fluid-solid interaction $\bm{F}_{\rm {f-s}}$:

\begin{equation}
\bm{F}_{\rm{f-s}}(\bm{x},t)=\kappa_{\rm w} \rho(\bm{x},t) \sum_{i=1}^N w_{i}s(\bm{x} +
\bm{v}_{i}\delta t)\bm{v}_{i},
\label{f-s:forces}
\end{equation}

where the strength of the fluid-solid interaction is reflected by $\kappa_{\rm w}$. The indicator function $s(\bm{x}+\bm{v}_{i}\delta t)$ in Eq. \ref{f-s:forces} is equal to one for solid nodes, but zero otherwise. $w_i$ are the weight coefficients \citep{ref13}. \\

In the simulation, the droplet ($D_0 = 2.3$ mm) was initially placed at a certain height above the fabric. As gravity effects could be neglected in the experiments, these were also not considered in the simulations. Both liquid and vapour phases were first initialized by imposing a zero impact velocity, after which the simulations were run for a short period of time to allow the liquid-vapour interface to reach equilibrium. Then, a uniform impact velocity $\rm V_i$ towards the surface was imposed on the liquid droplet.  The simulated fluid parameters used were: $\rho/\rho_c=3.06$ (liquid to critical density ratio), $\rho_v/\rho_c=0.028$ (vapour to critical density ratio) and the interfacial surface tension is $\sigma/p_c \hspace{0.05cm} D_0=0.0296$, corresponding to $\kappa=0.00468$ in Eq. \ \ref{eq:Pabtarget}. The critical density $\rho_c$ was computed at the critical temperature $T_c$ and critical pressure $p_c$ from the Peng-Rabinson equation of state \citep{ref16}. Previous studies have shown that a liquid to vapour density ratio of around 110 is more than sufficient to correctly capture the dynamics of droplets impacting on solid substrates \citep{ref17,sm2018}; these are used in this study. The dynamic viscosity $\eta$ of the liquid is set according to the Ohnesorge number $(Oh=\eta/\sqrt{(\rho\hspace{0.05cm} \sigma D_0 )})$ for the simulated water droplet. The pore size $B$ and spacing $L$ (Fig.\ \ref{fig:Schematic}a) are determined by keeping the aspect ratio of the droplet diameter to the pore size and pore spacing the same as in the experiments. The equilibrium solid-liquid contact angle in simulations is set to $\approx 70^\circ$, comparable to the contact angle of polyester \citep{Hsieh1998}. The size of the computational domain was determined using a grid independence study, giving a domain of $700\times700\times400$ grid nodes. Periodic boundary conditions were applied at the edges, where a wall boundary condition similar to \cite{ref13} was used for the top and bottom edges of the simulations as well as the solid surfaces.\\

Droplet impact was simulated for multiple surfaces similar to the surfaces used in the experiments. A smooth surface was used as a reference. The simulated patterned surface (Fig.\ref{fig:Schematic}b) had holes of the same dimensions as the pores of the 150-micron fabric. The fabric was recreated in the simulations by weaving rectangular `fibres' in a sinusoidal pattern, similar to the plain weave of the fabric used in the experiments. Droplet impact was simulated for both a fabric placed on a substrate and a fabric suspended in the air. For the former, droplet impact was simulated for cases where the fabric was attached to the substrate (no flow between the fabric and substrate), or placed above the substrate with a small gap of around 35 micron between the fabric and substrate, allowing for a flow of liquid between the two. For all simulations, the simulated `fibres' were considered to be rigid during droplet impact.

\section{Results and discussion} 
\begin{figure*}
\centering
\includegraphics[width=.8\textwidth]{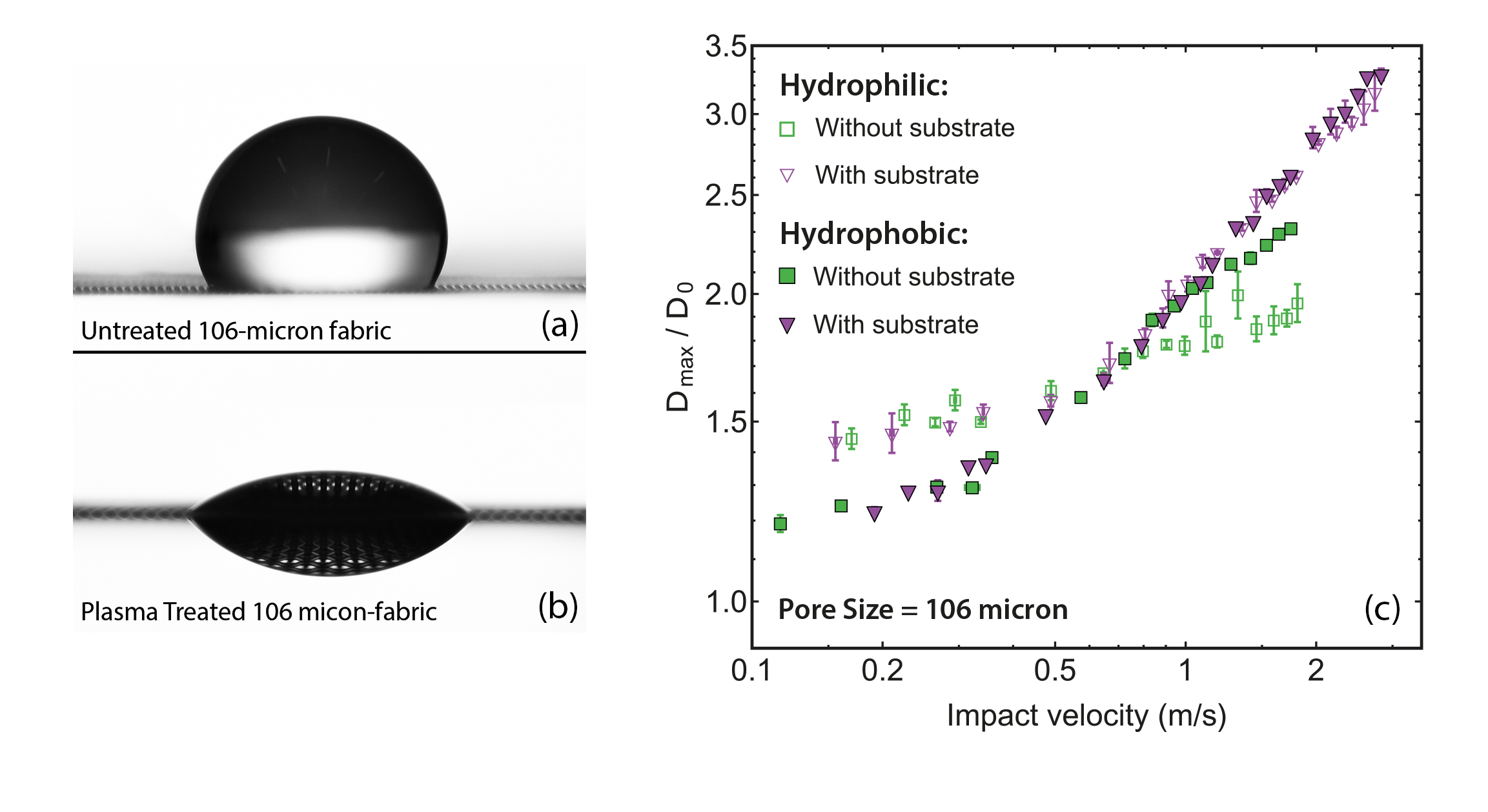}
\caption{Gently deposited water droplet at mechanical equilibrium on untreated (low wettability) (a) and plasma treated (high wettability) (b) 106-micron fabrics suspended in the air.  (c) Measured spreading ratio on plasma-treated (hydrophilic; open symbols) and untreated (hydrophobic; filled symbols) 106-micron fabrics. Fabrics were placed on a substrate (purple triangles) or spanned over a small gap of air (green squares). }
\label{fig:wettability}
\end{figure*}

\subsection{Droplet impact and fabric penetration}

Figure \ref{fig:highspeedimg}a shows high-speed images of a water droplet at maximum spreading on stainless steel (top row), fabrics with substrate (middle row) and fabric without a substrate (bottom row). Similar to earlier studies \citep{Kumar2018,Zhang2018}, the droplet penetrates through the fabric without a substrate when the impact velocity is high enough. Consequently, the measured droplet spreading ratio (Fig.\ref{fig:highspeedimg}b) of fabrics with and without substrate starts to deviate at the moment the droplet penetrates the fabric. The penetration velocity of the fabric is dependent on the pore size, where the penetration velocity of the 45 micron and 150 micron fabric found in this study are $1.5 \pm 0.1$ m/s and $0.8 \pm 0.1$ m/s, respectively.\\

A droplet can only penetrate a hydrophobic fabric (or mesh) when the dynamic pressure the droplet exerts on the fabric ($\sim \rho v^2$) is high enough to overcome the resisting capillary pressure ($\sim 4\sigma / B$) of the pores. Ryu \textit{et al.} showed that it is possible to calculate the penetration velocity $v_{p}$ at which a mesh is penetrated by balancing these pressures \citep{Ryu2017}:  

\begin{equation}
v_{p} = \sqrt{\frac{4 \sigma}{C_0 \rho B}}~,
\label{eq:vpen}
\end{equation}

where $C_0$ is a proportionality constant, which is equal to 2.78 for plain woven meshes with rectangular pores \citep{Ryu2017}. Using this equation, we find penetration velocities of around 1.52 m/s and 0.83 m/s for the 45 micron and 150 micron fabrics, respectively, agreeing very well with the penetration velocity determined from the drop impact measurements. Thus, the liquid penetration of the fabric indeed has a significant influence on droplet spreading spreading, in agreement with previous studies. \\

Fabric penetration by the droplet is also dependent on the wettability of the fabric \citep{Kota2012,Park2013,Hou2015}. The low wettability of the fibres leads to the capillary pressure inside the pore that resists the droplet pushing through the fabric. If the wettability of the fibres is increased, fabric penetration by the liquid is made easier as the capillary pressure disappears.\\ 

The effect of fabric wettability on the capillary pressure can be observed when a droplet is gently placed on the untreated fabric (Fig. {\ref{fig:wettability}}a) or plasma treated fabric (Fig. {\ref{fig:wettability}}b). For the untreated fabric, the capillary pressure resulting from the low fibre wettability keeps the droplet on top of the fabric, resulting in a high contact angle on the fabric ($116^\circ \pm 3^\circ$) droplet. When the wettability of the fibres is increased with the plasma treatment, the capillary pressure disappears and allows the droplet to pass through the pores and wet both sides of the fabric. The resulting contact angle of the top part of the droplet is significantly lower ($38^\circ \pm 3^\circ$) compared to the untreated fabric. Do note that these fabric contact angle are not equal to the contact angle of the polyester fibres, which could not be measured in this study. However, these images do show that the plasma treatment significantly changes the wettability of the fabric, which has a significant influence on how the droplet interacts with the fabric.\\

The change in wettability by the plasma treatment also influences droplet spreading on top of the fabric. First, increasing the wettability of the fabric increases the droplet spreading ratio at low impact velocities, both for the fabrics with and without substrates (Fig. \ref{fig:wettability}c). This is a similar effect that is observed when the wettability is increased for smooth surfaces \citep{deGoede2019}. The second wettability effect on droplet spreading is that it reduces the droplet spreading ratio on plasma-treated fabrics without substrate (open green squares in Fig. \ref{fig:wettability}) compared to the droplet spreading ratio on untreated fabrics without substrate (filled green squares). Comparing high-speed videos of the untreated fabric and plasma treated fabric (Movies 1 and 2 in the Supplementary Material, respectively) reveals that the increased wettability results in more liquid being pushed through the fabric. It is likely that the larger amount of liquid passing through the fabric reduces the droplet spreading ratio as less volume is left on top of the fabric to spread outwards.\\
\begin{figure*}
\centering
\includegraphics[width=\textwidth]{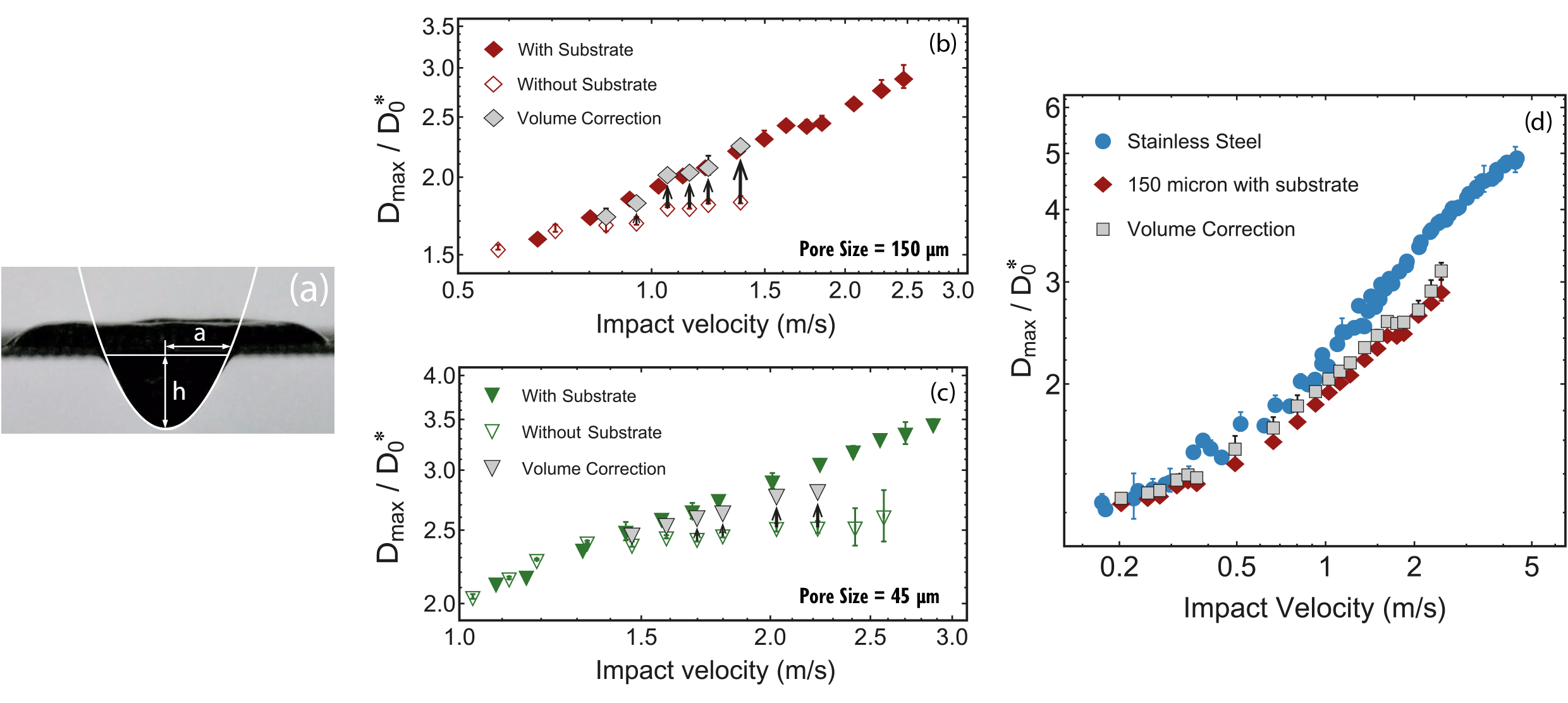}
\caption{(a) Photograph of liquid penetrating a fabric without substrate with the white line a paraboloid of base radius $a$ and height $h$. (b,c) Spreading ratios as a function of impact velocity for 150 (b) and 45 (c) micron pore-size fabrics with and without substrate, with the latter values also corrected for liquid loss due to fabric penetration using Eq. \eqref{eq:volcorr} (grey symbols). (d) Spreading ratios for a smooth reference surface (blue symbols) and a 150-micron fabric with substrate, both without (red diamonds) and with (grey squares) correction for the volume lost inside the pores.}
\label{fig:voladjust}
\end{figure*}

\begin{figure*}
\centering
\includegraphics[width= .9\textwidth]{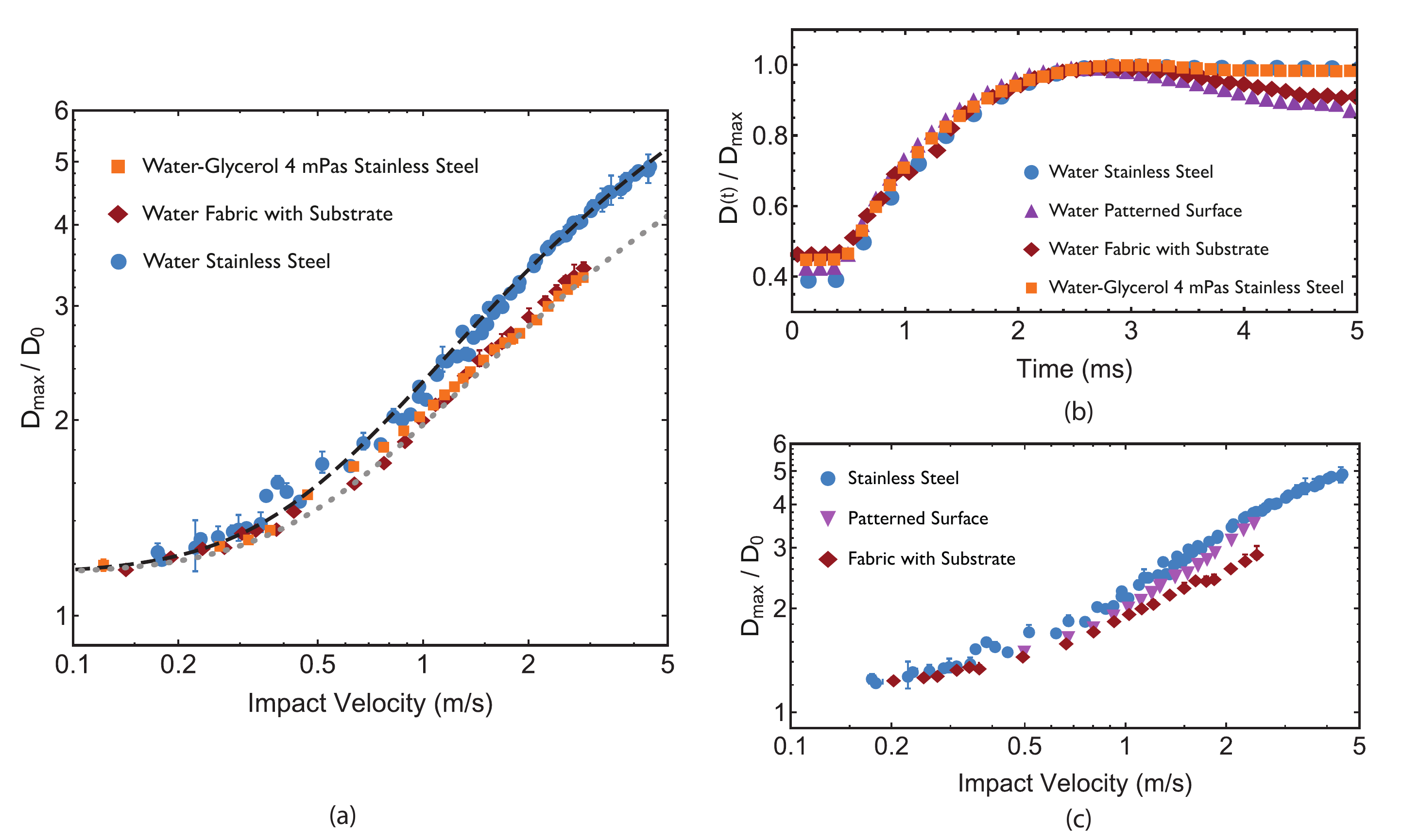}
\caption{(a) Measured spreading ratio for water impacting a stainless steel surface (blue circles) and a 150 $\mu$m pore size fabric with substrate (red diamonds) compared to the measured spreading ratio of a 1:1 water-glycerol mixture (4 times the viscosity of water) impacting stainless steel. The dashed and dotted lines are the best fits of the theoretical predictions of Lee \emph{et al.} for water and water glycerol mixture impacting stainless steel, respectively. (b) Measured droplet diameter at $v \approx 1.25$ m/s as a function of time from the moment of impact ($t = 0$) for the surfaces shown in (a) and (c), rescaled with $D_{max}$. (c) Measured droplet spreading ratio on stainless steel (blue circles), patterned surface (purple triangles) and 150 micron fabric with substrate.}
\label{fig:watglyimpact}
\end{figure*}

\begin{figure}
\centering
\includegraphics[width=0.5\textwidth]{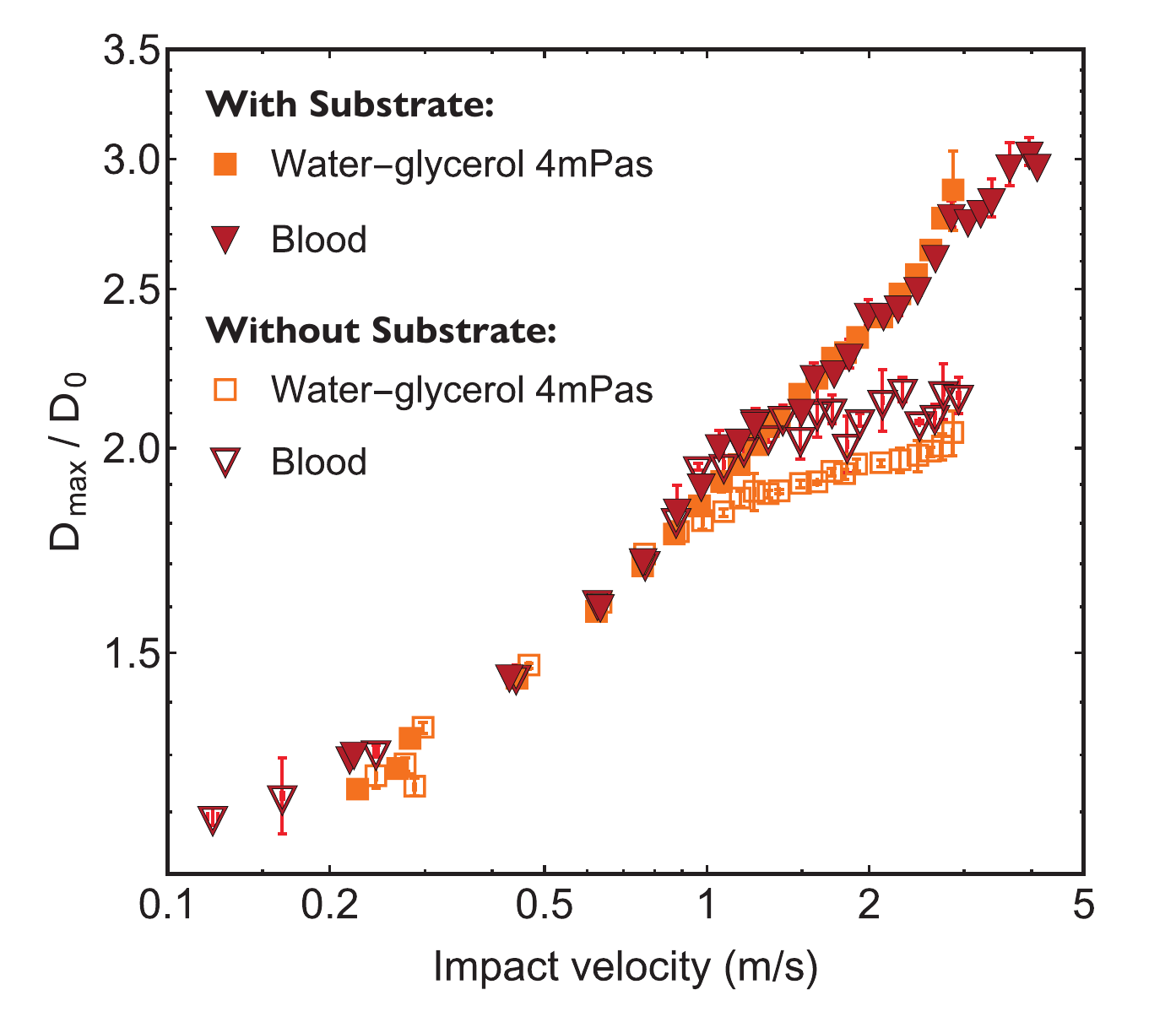}
\caption{Measured spreading ratio of a 1:1 water-glycerol mixture ($\eta= 4$ mPas; orange squares) and blood ($\eta_{\infty}= 4.8$ mPas; red triangles) on a 150-micron fabric with and without substrate (filled and open symbols, respectively).}
\label{fig:blood}
\end{figure}

\begin{figure*}
\centering
\includegraphics[width=.8\textwidth]{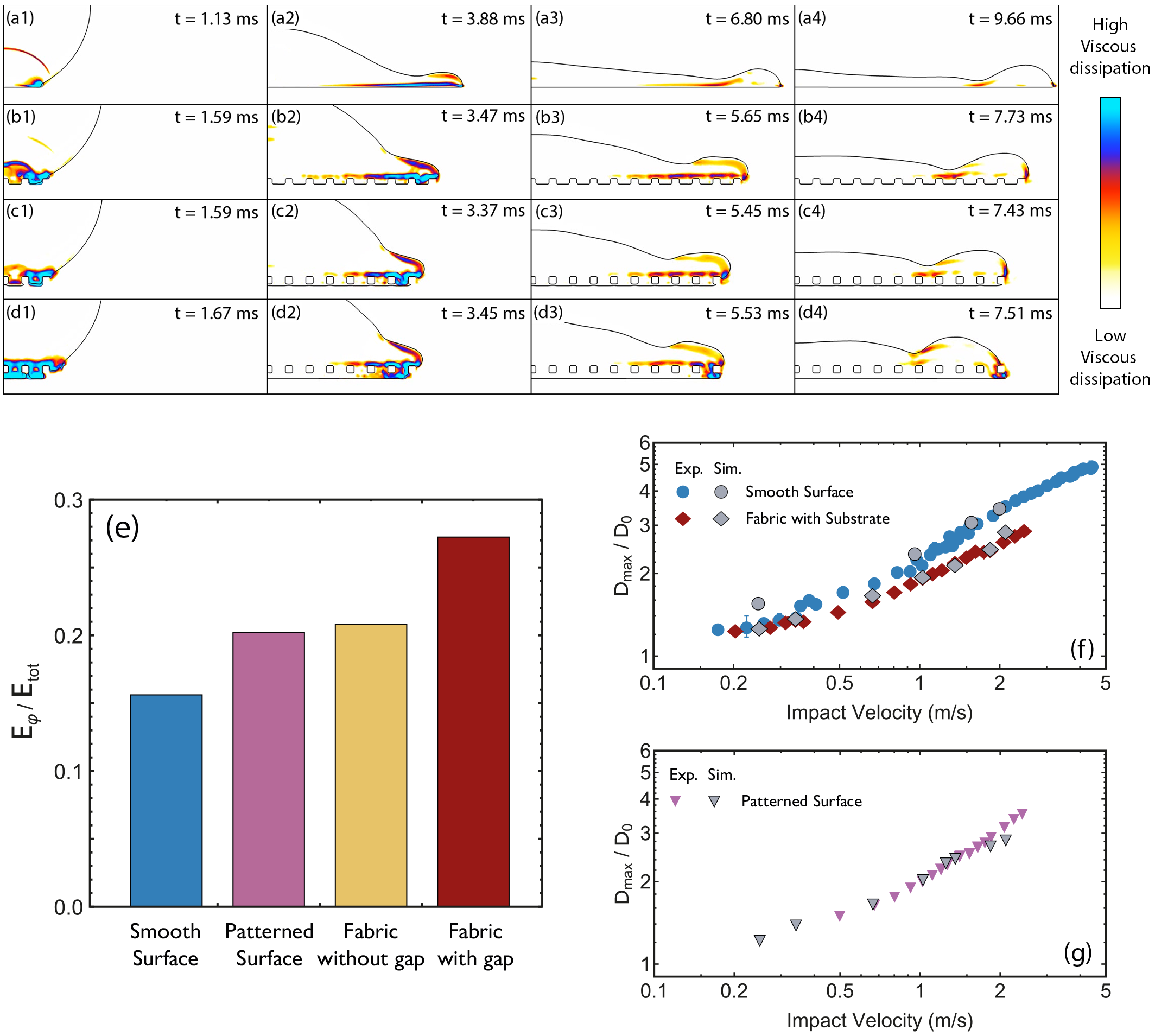}
\caption{Simulation results. (a-d) Cross sections of droplets impacting on a smooth surface (a1-4), a patterned surface (b1-4), fabric on a substrate without a gap (c1-4) and with a gap (d1-4) at four subsequent times from the moment of impact to maximum spreading (from left to right). The impact velocity of each droplet is equal to 1.25 m/s. The colour scale qualitatively depicts the amount of viscous dissipation inside the droplet, from white (low viscous dissipation) to blue (high viscous dissipation). (e) Ratio of the total dissipated energy with the total energy of the system of a spreading droplet on a smooth surface (blue bar), patterned surface (purple) and fabric without (yellow) and with a gap between the fabric and substrate (red). The total viscous dissipation is determined from the moment of impact up to maximum spreading. (f) Comparison between the measured spreading ratio determined from experiments (coloured symbols) and fabric with gap simulations (grey symbols) for the smooth surface (circles) and 150-micron fabric (diamonds) (g) Comparison of the experiments and simulations of droplet spreading on a patterned surface.}
\label{fig:vissim}
\end{figure*}  

\begin{figure*}
\centering
\includegraphics[width=\textwidth]{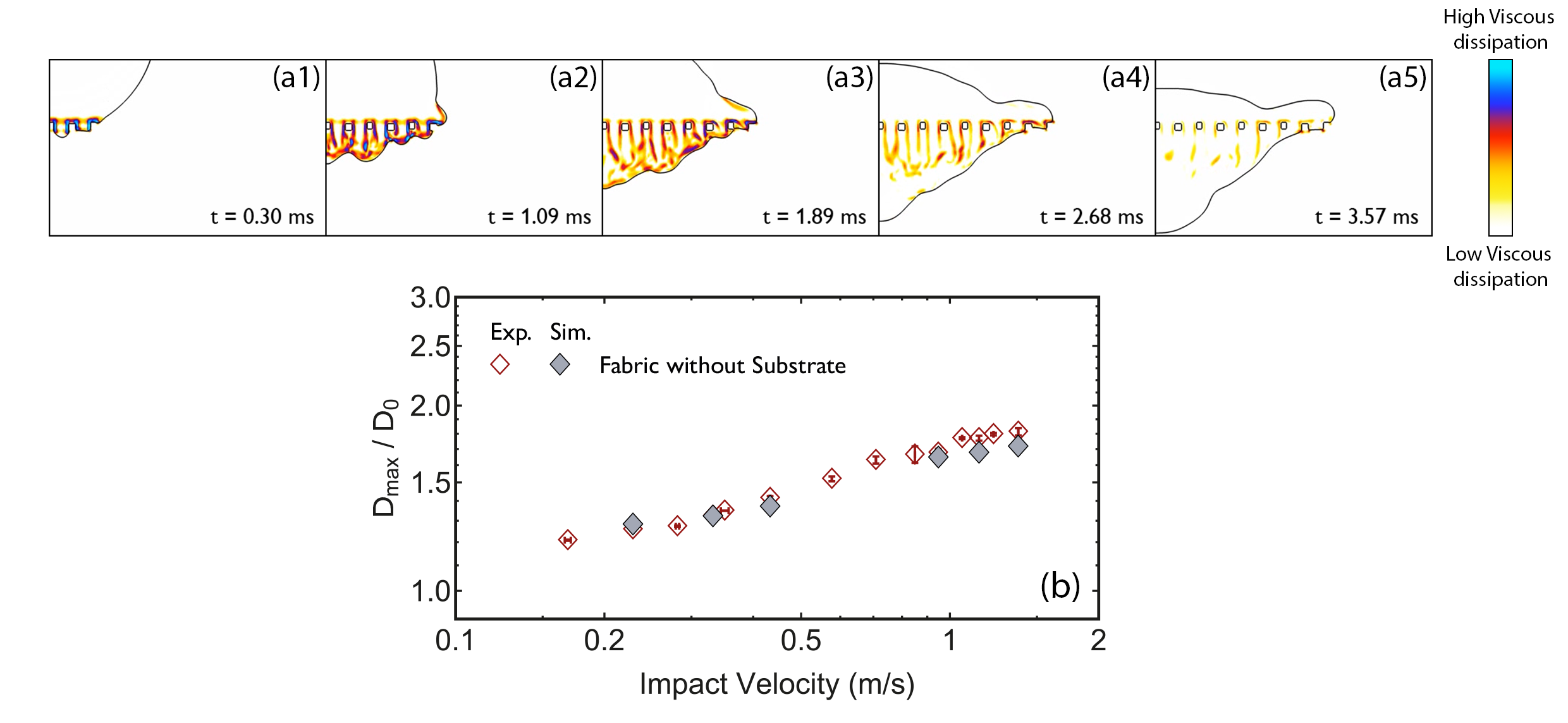}
\caption{(a) Viscous dissipation inside a droplet impacting ($v = 1.4$ m/s) on a fabric without substrate shown from the moment of impact ($t = 0$) to maximum spreading. The colour scale gives a qualitative depiction of the viscous dissipation from white (low viscous dissipation) to blue (high viscous dissipation). (b) Comparison between experiments (red diamonds) and simulations (black diamonds) of water droplet spreading on fabrics without substrate.}
\label{fig:nosub}
\end{figure*}

A reduction in droplet volume on top of the fabric influences the spreading difference between fabrics with and without substrate underneath. Due to the volume reduction, the initial diameter $D_0$ overestimates the actual liquid volume spreading on top of the fabric, decreasing the droplet spreading ratio. To correct for this volume loss, we experimentally estimate the liquid volume that penetrates the (untreated) fabric. Assuming that the liquid underneath the fabric takes the shape of a paraboloid (Fig. \ref{fig:voladjust}a), the volume of the penetrated liquid $V_{pen}$ can be estimated from the images by determining the base radius $a$ and height $h$ of the paraboloid:
\begin{align}
V_{pen} = \frac{\pi}{2} a^2 h
\end{align}   

By subtracting $V_{pen}$ from the volume of the droplet, an adjusted initial diameter $D_0^*$ can be determined, and with it the volume-corrected spreading ratio:

\begin{align}
\frac{D_{max}}{D_0^*} = \frac{D_{max}}{\sqrt[\leftroot{-1}\uproot{2}\scriptstyle 3]{D_0^3 - a^2 \hspace{.1cm} h}}
\label{eq:volcorr}
\end{align}

Using the above equation, the volume loss corrected spreading ratio (grey symbols; Fig. \ref{fig:voladjust}b and \ref{fig:voladjust}c) on both the 150 micron (Fig. \ref{fig:voladjust}b) and 45 micron (Fig. \ref{fig:voladjust}c) fabrics without substrate is determined for every drop impact measurement above the penetration velocity. The volume correction for both fabrics works very well, as the spreading ratios of the fabric with and without substrate become comparable after the volume correction is applied. For the 45-micron fabric, we still observe a slightly lower spreading ratio for fabrics without substrate after the volume correction. This is likely due to the liquid below the fabric attaining an irregular shape when it coalesces underneath the fabric, making the volume estimation with a paraboloid less accurate for the 45-micron fabric. Nevertheless, these results show that the difference in spreading for fabric with and without substrate is fully determined by the loss of liquid when the droplet penetrates the fabric. \\

If the droplet impacts a fabric without a substrate fast enough, the penetrating liquid moves too fast to coalesce underneath the fabric and liquid remains separated as liquid `fingers' instead. These fingers subsequently break up into many droplets, resulting in a spray where the droplets that are on the order of 100 microns in size (Fig. \ref{fig:highspeedimg}a; bottom right). Droplet fragmentation below a mesh in itself is a fascinating phenomenon, attracting interest due to several recent studies as it could be used as a novel method to create sprays \citep{Brunet2009,Soto2018}, although it was recently shown that it currently performs poorly compared to other atomisation methods \citep{Kooij2019}. As the main focus of this study lies on droplet spreading on top of the fabrics however, droplet fragmentation is not discussed here.\\

Being able to correct for the volume loss during droplet spreading on fabrics without substrate, we can now compare droplet spreading on fabrics with droplet spreading on smooth surfaces (Fig. \ref{fig:highspeedimg}b). At low impact velocities ($v < 1$ m/s), no significant difference can be observed between droplets spreading on fabrics with substrate and smooth surfaces. For high velocity droplet impacts ($v > 1$ m/s) however, the droplet spreading ratio on the fabrics with substrates is significantly lower compared to the droplet spreading ratio on the smooth surface. This deviation increases with increasing pore size. \\ 

We next investigate whether this difference in spreading is also due to a loss of liquid volume, as part of the droplet is used to fill up the pores of the fabric. With the assumptions that the pores are rectangular (with volume of $B^2 L$) and that all pores underneath the droplet at maximum spreading are filled, it is possible to estimate the total droplet volume loss on the fabric with substrate at maximum spreading, assuming there is no gap between the fabric and substrate:

\begin{equation}
V_{pen} = N_p V_p = \frac{\pi}{4}D_{max} \frac{B^2 L}{\left( B + L \right)^2} ,
\label{eq:volcorrsub}
\end{equation}
where $N_p $ is the number of filled pores and $V_p$ is the volume of a single pore. Correcting the volume loss using Eqs \eqref{eq:volcorr} and \eqref{eq:volcorrsub} shows that the lost volume is too small to account for the difference in spreading between a fabric with substrate and smooth surfaces. (Fig. \ref{fig:voladjust}d). Thus, the difference in the droplet spreading ratio between the smooth surfaces and fabrics has to be due to a different mechanism entirely.
\\

\subsection{Viscous dissipation}

As mentioned earlier, the kinetic energy of an impacting droplet during spreading is either transformed into surface energy or dissipated by the viscous forces inside the droplet. Eggers \emph{et al.}, Laan \emph{et al.} and Lee \emph{et al.} found a relation between the droplet spreading ratio and the impact velocity and fluid properties by interpolating between the capillary regime ($\propto \mathrm{We}^{1/2}$) and viscous regime ($\propto \mathrm{Re}^{1/5}$) using a first-order Pad\'e approximant \citep{Eggers2010,Laan2014}, which was modified by \cite{Lee2015} to account for low impact velocity droplet spreading:

\begin{equation}
\left(\frac{D_{max}}{D_0}\right)^2 -\beta_0^2 = \frac{\mathrm{We}^{1/2}}{7.6 + \mathrm{We}^{1/2}}\hspace{0.1cm} \mathrm{Re}^{1/5} .
\label{eq:spreadingmodel}
\end{equation}

Here, $\beta_0$ is defined as the value of the maximum spreading ratio at zero impact velocity, which is dependent on the liquid surface tension and surface wettability \citep{Lee2015,deGoede2019}. The above equation shows very good agreement with our experimental data for water droplets impacting the smooth surface (black line in Fig. \ref{fig:highspeedimg}b, where $\beta_0$ was used as a fitting parameter). Lee \textit{et al.} also showed that the droplet spreading ratio at high impact velocities decreases when the viscosity of the liquid is increased, as viscous losses become more important at higher impact velocities. We investigate the role of liquid viscosity by comparing the droplet spreading ratio of a water-glycerol mixture impacting a smooth surface ($\eta = 4$ mPas; orange squares in Fig. \ref{fig:watglyimpact}a) to that of a water droplet impacting a 150-micron fabric (Fig. \ref{fig:watglyimpact}a). Interestingly, the spreading curves are similar. These measurements suggest that changing the smooth surface to a fabric has an equivalent effect on droplet spreading as increasing the viscosity of the fluid does. The best fit of Eq. \eqref{eq:spreadingmodel} for the water-glycerol mixture also predicts the measured spreading ratio on fabrics well, allowing to determine the the spreading dynamics of water on a fabric substrate by using an `effective' viscosity higher than the viscosity of the liquid. Furthermore, the spreading dynamics of a droplet between the moment of impact and maximum spreading for all liquids and surfaces studied here are similar when the droplet spreading diameter $D(t)$ is rescaled with the maximum spreading diameter (Fig. \ref{fig:watglyimpact}a). A droplet reaches maximum spreading at roughly $2.8 \pm 0.2$ ms after impact, irregardless of its viscosity or the surface it spreads over. Thus, any changes in the droplet spreading ratio are fully governed by the maximum spreading diameter of the droplet. We hypothesise that the smaller droplet spreading ratio on fabrics are due to an increase in viscous losses inside the droplet when it spreads over the fabric.\\

A likely candidate for the increased viscous dissipation inside a droplet are the pores of the fabric: when a droplet spreads over a surface, liquid enters the pores underneath until they are full, where the liquid in the pore comes to a full stop and the kinetic energy is dissipated by the viscous forces. If that would be true, droplet spreading on fabrics with substrate should be similar to droplet spreading on a patterned surface. We find that this is not the case (Fig. \ref{fig:watglyimpact}c). Although droplet spreading on the patterned surface and 150 micron fabric are similar at first, the droplet spreading ratio of the patterned surface diverges from that of the fabric at high impact velocities. 
At high impact velocities, the droplet spreading ratio on the patterned surface becomes comparable to the droplet spreading ratio on smooth surfaces, which could be caused by the droplet spreading too fast for the liquid to move into the pores and push the air out, resulting the droplet to spread over the air pockets instead of filling the pores, limiting the influence of the pores on droplet spreading. However, whether this is true cannot be concluded with the results presented here. Interestingly however, the droplet spreading ratios of both fabric with substrate and patterned surface starts to deviate at the moment the impact velocity is higher than the penetration velocity of the 150 micron fabric ($0.8 \pm 0.1$ m/s). We thus propose that the substrate underneath the fabric blocks the downward flow of the fluid and redirects the fluid in between the fabric and substrate due to the pores of the fabric being connected. So not only does the droplet loses energy due to the pores of the fabric, the viscous losses that could be caused by the droplet pushing itself between the fabric and substrate also have to be taken into account.\\

\subsection{Blood droplet impact}

Before discussing the viscous losses inside droplets impacting a fabric mesh, droplet spreading of blood is discussed first. If the shear thinning properties of blood are indeed important for droplet spreading on fabrics, these properties should be taken into account for the viscous losses inside a droplet. Comparing the spreading ratio of blood (Fig. \ref{fig:blood}, red triangles) and the water-glycerol mixture (orange squares), which has a viscosity similar to the high shear rate viscosity $\eta_{\infty}$ of blood, reveals two interesting features. First, the spreading ratios on fabric as a function of impact velocity of blood and the water-glycerol mixture (red triangles and orange square, respectively) are roughly similar. This observation is in line with that of \cite{Laan2014} for the same liquids spreading on a smooth surface. Secondly, the penetration velocity, which is the impact velocity at which the spreading curves with and without a surface underneath the fabric start deviating from each other, is different for the two liquids. For the water-glycerol mixture, it is around 1 m/s (higher than the penetration velocity of water) and for blood 1.5 m/s.\\

This suggests that blood penetrates the fabric less than its Newtonian counterpart. When a liquid pushes through a constriction, not only are the shear stresses important but also the elongational stresses applied on the liquids. While blood shows non-Newtonian viscous behaviour when it is sheared (e.g. a spreading droplet), studies have shown that it also exhibits viscoelastic behaviour when subjected to elongational stresses \citep{Brust2013,Campo2013,Kar2018}. This viscoelastic behaviour in the extensional flow could cause more resistance against fabric penetration, leading to the increased penetration velocity observed in this study. As the viscoelastic behaviour in the extensional flow of blood only results in a decrease in the amount of liquid pushing through the fabric, it has no influence on the droplet spreading ratio after the spreading ratio is corrected for the volume loss due to the liquid penetration of the fabric. Thus we conclude that blood not only spreads like a Newtonian fluid on smooth surfaces \citep{Laan2014}, but on fabrics as well. 

\subsection{Droplet impact simulations}

To determine whether the viscous dissipation inside a spreading droplet is higher when spreading over a fabric, the viscous losses inside the droplet during spreading were determined using an entropic lattice Boltzmann simulation method. With these simulations, the liquid flow velocity $v$ can be calculated inside the droplet during spreading, which subsequently can be used to determine the dissipation function for each simulation time step $\Phi$ inside the droplet by calculating the shear rate in each grid node $i,j$:

\begin{equation}
\Phi = \frac{\mu}{2} \left( \frac{\partial v_i}{\partial x_j } + \frac{\partial v_j}{\partial x_i} \right)^2
\end{equation}

The viscous losses inside the droplet during spreading on each simulated surface are shown in Fig. \ref{fig:vissim}. For smooth surfaces, the majority of the viscous dissipation takes place at the interface between the spreading droplet and surface (panels a1-a4). This is expected as the surface generates a significant shear stress inside the liquid during spreading. For the patterned surface (b1-b4) and the fabric attached to the substrate (c1-c4), the viscous dissipation on top of the surface is similar to that of the smooth surface, but there are additional viscous losses inside the pores, caused by the liquid flow filling the pores until they are filled completely. When there is a gap between the fabric and substrate (d1-d4), the liquid is pushed between the fabric and substrate by the solid substrate, leading to viscous losses the moment the liquid pushes itself between the fibres of the fabric and the substrate.\\

For each simulated surface, the total energy lost due to viscous dissipation $E_\phi$ inside the spreading droplet is determined by summing all viscous losses inside the droplet from the moment of impact up to the moment the droplet reaches maximum spreading. $E_{\phi}$ is then normalised with the droplet's total energy $E_{tot}$ (Fig. \ref{fig:vissim}e). On a smooth surface (blue bar), a water droplet loses around 16 percent of its kinetic energy due to viscous forces during impact. The viscous losses for the patterned surface (purple bar) are significantly higher, indicating that the flow inside the pores indeed leads to an increase of viscous losses. Furthermore, the similar viscous dissipation for the patterned surface and fabric that is attached to the surface (no connection between the pores; yellow bar) suggests that the additional surface roughness caused by the weaving of the fabric has no significant influence on the viscous dissipation inside the droplet and hence on the process of droplet spreading. However, as we only considered a single type of fabric weaving in this study, no definite conclusion can be given. Finally, if the fabric is detached from the substrate, and liquid is allowed to flow in between the fabric and substrate, the viscous dissipation inside the droplet (red bar) significantly increases compared to the patterned surface.\\

To determine whether the lower spreading ratios on fabrics are indeed caused by the extra viscous losses inside the droplet, the droplet spreading ratio were determined for the simulated smooth surface, patterned surface and fabric with a gap. We find that the simulated droplet spreading ratio (Fig. \ref{fig:vissim}f) and the experimental measurements (Fig. \ref{fig:vissim}g) agree very well. For the smooth surface, the simulated spreading ratio is slightly higher in the low impact velocity regime. However, the contact angle of the liquid on the simulated smooth surface was lower ($\theta = 70^\circ$) compared to that of the stainless steel surface ($\theta = 80^\circ$) used in the experiment. Low impact velocity spreading is dependent on the wettability of the surface \citep{deGoede2019}, and thus the higher spreading ratio for the simulation is most likely be due to the lower contact angle used in the simulations. 

The experimental droplet spreading ratio of the patterned surface is described well by the simulations of a smooth patterned surface (Fig. \ref{fig:vissim}g) as well. The experimental droplet spreading ratio on fabric with a substrate only agrees well with the simulations for the fabric with a gap (Fig. \ref{fig:vissim}f), indicating that the viscous dissipation inside the flow between fabric and substrate is indeed important during droplet impact on fabrics. The simulated droplet spreading ratios for both the patterned surface and fabric with a gap seem to deviate from experiments at the highest simulated impact velocity. However, this deviation is relatively small and is probably caused by errors due to being at the impact velocity limit that can be simulated with the simulation method used here. Also note that the simulations only considers interactions between the liquid and its vapor, but not the interactions between the liquid and any surrounding gas. It is thus possible that the deviation between the simulations and experiment for the patterned surface could also originate from the droplet spreading over the air pockets inside the pores of the patterned surface in the experiments, which cannot be simulated with the used simulation method. However, as these deviations between experiments and simulations happen at the maximum impact velocity limit of the simulation method, no definite conclusion cannot be given. However, the good agreement between the measured and simulated droplet spreading ratios confirms that the difference in spreading on smooth surfaces and fabrics is indeed due to an increase in viscous losses, which is caused by both the flow into the fabrics pores and the flow in between the fabric and substrate.\\

The increase in viscous dissipation during droplet impact is not only observed for fabrics with substrate, but also for fabrics without substrates underneath. For droplet impact simulations on fabrics without substrate (Fig. \ref{fig:nosub}a), the extra viscous dissipation originates from the downward flow of the fluid. The liquid pushes itself through the pores in columns that coalesce, leading to the extra viscous dissipation underneath the fabric. The viscous losses inside a droplet pushing through the  fabric without a substrate has a significantly higher viscous dissipation ($E_\varphi/E_{tot} = 0.43$) compared to that of a droplet spreading on a smooth surface ($E_\varphi/E_{tot} = 0.16$). The viscous dissipation for the fabric without substrate is also higher compared to the viscous losses for the fabric with a gap ($E_\varphi/E_{tot} = 0.27$), although this can be partly ascribed to the impact velocity being higher for the droplet impact simulation of the fabric without substrate ($v = 1.4$ m/s) compared to that of the simulations on the fabric with substrate ($v = 1.25$ m/s). The spreading ratios obtained from the simulations are slightly lower than those of the experiments (Fig. \ref{fig:nosub}b) but still agree well. These measurements show that the extra viscous losses also occur inside the droplet during spreading over fabrics without a substrate.\\

\section{Conclusion}

In this study, we investigated the influence of the fabric penetration by the liquid on droplet spreading on monofilament polyester fabrics. Using high-speed imaging, we showed that the droplet spreading ratio is influenced by the penetration of the fabric. By applying a volume correction on the droplet spreading ratio on fabrics without substrate, we show that the difference in spreading between fabrics with and without substrate is caused by fabric penetration in the form of liquid volume loss, in agreement with earlier studies. By comparing experiments with entropic lattice Boltzmann simulations, we show that the lower droplet spreading ratio on fabrics at high impact velocities is due to increased viscous losses inside the droplet, of which a significant part originates from the droplet pushing itself through the fabric or in between the fabric and substrate during droplet spreading. Finally, although there is a difference in the penetration dynamics for blood, we show that blood can still be approximated as a Newtonian fluid during droplet spreading on fabrics. \\

Our study shows that droplet spreading is significantly influenced by the fabric geometry for even the most simple of fabrics. For applications such as ink-jet printing on textiles, these results could be important: they show that an ink droplet needs to hit the fabric at a higher impact velocity compared to smooth surfaces to cover the same area of fabric. However, our results also show that if the impact velocity becomes too large, the ink can push itself through the fabric and spread out in between the fabric and substrate as well, which can have undesirable effects on the other side of the fabric. Our results also show that, although it is currently not possible to fully determine the total viscous dissipation during droplet spreading on fabrics beforehand, the effect on droplet spreading can be accounted for in the pre-existing spreading model on smooth surfaces with an`effective' viscosity that is higher than the actual viscosity of the fluid. The actual value of this effective viscosity is dependent on the fabric geometry and the liquid that spreads over the fabric. Finally, our conclusion that blood spreads similar to a Newtonian fluid on these simple fabrics implies it is still possible to use the spreading model to find a relation between the size of the bloodstain and the impact velocity of the droplet spreading on a fabric, something that was only currently possible for blood droplets spreading on smooth surfaces.   

\section*{Acknowledgements}

A.M.M., D.D., and J.C. acknowledge the support by the Swiss National Science Foundation (Project no. 200021\_175793). The computational resources were provided by the Swiss National Supercomputing Center (CSCS) under project number s823. 

\newpage
\bibliographystyle{naturemag}
\bibliography{fabricspreading}
\end{document}